\documentclass[aps,twocolumn,showpacs,superscriptaddress,groupedaddress]{revtex4-1}  

\usepackage{graphicx}  % needed for figures
\usepackage{dcolumn}   % needed for some tables

\usepackage{bm}        % for math
\usepackage{amssymb}  
\usepackage{amsmath}
\usepackage{eqnarray,amsmath}% for math
\usepackage{lipsum} % j
\usepackage[colorinlistoftodos]{todonotes}
\usepackage[colorlinks=true, allcolors=blue]{hyperref}
\usepackage{physics}
\usepackage{float}
\usepackage{hyperref}
\usepackage{braket}
\usepackage{subcaption}
\begin{document}

	\global\long\def\dg{^{\dagger}}
	\global\long\def\sm{\sum_{<i,j>}}
	\global\long\def\up{\uparrow}
	\global\long\def\dn{\downarrow}
	\global\long\def\ri{\mathbf{r_{i}}}
	\global\long\def\rj{\mathbf{r_{j}}}
	\global\long\def\sg{\sigma}
	\global\long\def\si{\mathbf{S_{i}}}
	\global\long\def\sj{\mathbf{S_{j}}}
	\global\long\def\pp{\mathcal{P}_{0}}
	\global\long\def\qq{\mathcal{Q}_{0}}
	\global\long\def\qi{\frac{1}{\qq H_{0}\qq-E_{i}}}
	\global\long\def\qj{\frac{1}{\qq H_{0}\qq-E_{j}}}
	\global\long\def\qk{\frac{1}{\qq H_{0}\qq-E_{k}}}
	\global\long\def\jo{J_{n}(\zeta_{12})}
	\global\long\def\ji{J_{n+1}(\zeta_{12})}
	\global\long\def\js{J_{n+2}(\zeta_{12})}
	\global\long\def\nu{\Ket{0}}
\title{Orbital Floquet Engineering of Exchange Interactions in Magnetic Materials }	
\author{Swati Chaudhary}
\email{swatich@caltech.edu}  
\author{David Hsieh}
\author{Gil Refael}

\affiliation{Institute of Quantum Information and Matter, California Institute of Technology, Pasadena, California 91125, USA}

\begin{abstract}
We present a new scheme to control the spin exchange interactions between two magnetic ions by manipulating the orbital degrees of freedom using a periodic drive. We discuss two different protocols for orbital Floquet engineering. In one case, we modify the properties of the ligand orbitals which mediate magnetic interactions between two transition metal ions. While in the other case, we mix the $d$ orbitals on each magnetic ion. In contrast to previous works on Floquet engineering of magnetic properties, the present scheme makes use of the AC stark shift of the states involved in the exchange process. 
%Exchange interactions, orbitals, virtual energy modification
\end{abstract}
\maketitle
%\section{Introduction}

%Controlling the properties of the materials to optimize their performance and to realize novel phases of matter has always been at the core of the condensed matter physics[CHANGE].   Floquet engineering has the potential to manipulate the properties of quantum many body systems. 
Periodic drive is emerging as an intriguing  tool for controling and manipulating different quantum many-body systems. The evolution of periodically driven systems  can be described by an effective time-independent Floquet hamiltonian~\cite{PhysRev.138.B979}, which depends on the drive parameters. Floquet engineering has been invoked in contexts ranging from the generation of artificial gauge fields  to realization of many-body localization \cite{RevModPhys.89.011004,F_PhysRevB.79.081406,F_PhysRevB.84.235108,F_PhysRevLett.105.017401,F_lindner2011floquet,F_PhysRevX.4.031027,F_PhysRevX.6.021013,F_PhysRevLett.110.026603,F2_PhysRevB.96.020201,F_PhysRevLett.110.026603,F_fausti2011light,F_PhysRevLett.116.176401,F_PhysRevLett.112.156801,F2_PhysRevLett.109.145301,F2_PhysRevLett.118.115301,F2_baum2018dynamical,F2_goldman2014light,F2_holthaus2015floquet,F2_itin2015effective,FK_PhysRevB.97.085405,FK_PhysRevB.95.104308,F_PhysRevB.95.045102,FE_PhysRevB.96.155435,FM_PhysRevB.95.134508,FM_PhysRevB.95.155407,FM_PhysRevB.90.205127,FM_PhysRevB.96.155438,FM_yang2018floquet,FM_PhysRevB.95.174306,FSL_jotzu2014experimental,FQ_PhysRevB.97.035123,MagFl_owerre2018photoinduced,F0_PhysRevB.96.205127,F0_PhysRevB.97.075420,F0_PhysRevLett.119.267701,F0_inoue2018floquet,FKondo_PhysRevB.96.115120,Fweyl_PhysRevB.94.235137,FWeyl_PhysRevB.94.121106,FWeyl_fu2017phase,Fweyl_niu2018tunable,Fweyl_PhysRevB.96.041205,FWeyl_PhysRevB.96.041206,Fweyl_PhysRevB.96.041126,Fweyl_PhysRevB.95.144311,Fweyl2_PhysRevB.96.041205,Fweyl_PhysRevLett.117.087402} with ultracold atoms in optical lattices. These methods can potentially provide an external control knob for material properties, and can be naturally applied to controlling quantum materials~\cite{basov2017towards,oka2018floquet}.

  Recent works~\cite{Mentink2014,mentink2015ultrafast,mentink2017manipulating,Bukov2016,hejazi1PhysRevLett.121.107201,hejazi2018floquet} have discussed how  Floquet engineering  can be used to manipulate the exchange interactions in extended antiferromagnetic (AFM) Mott insulators. These modifications can be understood in terms of the properties of the Floquet hamiltonian that arise from photo-assisted hopping. They feature a renormalized electronic hopping, and, therefore, also a renormalized energy splittings in the effective Floquet Hamiltonian. These works assume direct hopping between two magnetic ions, and we refer to them as photo-modified direct hopping scheme henceforth.
 
 In transition metal (TM) compounds, ligand ions play a crucial role in spin exchange processes. For example, in 2D  transition metal trichalcogenides(TMTCs), the magnetic interactions are mainly mediated by ligand ions which provide multiple channels for spin exchange. The effects of a periodic drive are sensitive to these exchange pathways~\footnote{In preparation}. These ligand ions make an accurate description of the system under a periodic drive harder, but at the same time, the extra degrees of freedom such as orbitals of these ions can be manipulated to modify the exchange interactions. The magnetic coupling induced via ligand ions depends on the electronic energy and the shape of the orbitals available for spin exchange. Furthermore, the strong orbital-spin interplay   of TM ions also affects their electronic and magnetic properties~\cite{goodenough1963magnetism,khomskii1997interplay,khomskii2005role,tokura2000orbital,kugel1973crystal,streltsov2017orbital}. Many previous works have successfully manipulated some orbital properties using strain~\cite{PhysRevB.90.045128} and heterostructuring~\cite{PhysRevB.95.205131,PhysRevLett.114.026801}. 
 %also add pressurization
 %hybridization via AC Stark effect
 
In this manuscript, we explore the possibility of modifying the exchange interactions by manipulating the orbital degrees of freedom with a periodic drive. We propose an alternative scheme for Floquet engineering of exchange interactions. This scheme results in significant changes in the magnetic coupling strength even at Electric fields smaller than the one required for the photo-modified direct hopping scheme. 
We explore this scheme by using a toy model where strong time-dependent electric field couples two orbitals of the ligand ion and alters the exchange interactions significantly. We also explore the implications of a similar scheme involving the hybridization of orbitals on TM ions. Furthermore, we discuss the possibility of using a phonon drive to control the exchange interactions.% in some AFM Mott insulators.
\begin{figure} [H]
	\includegraphics[scale=0.312]{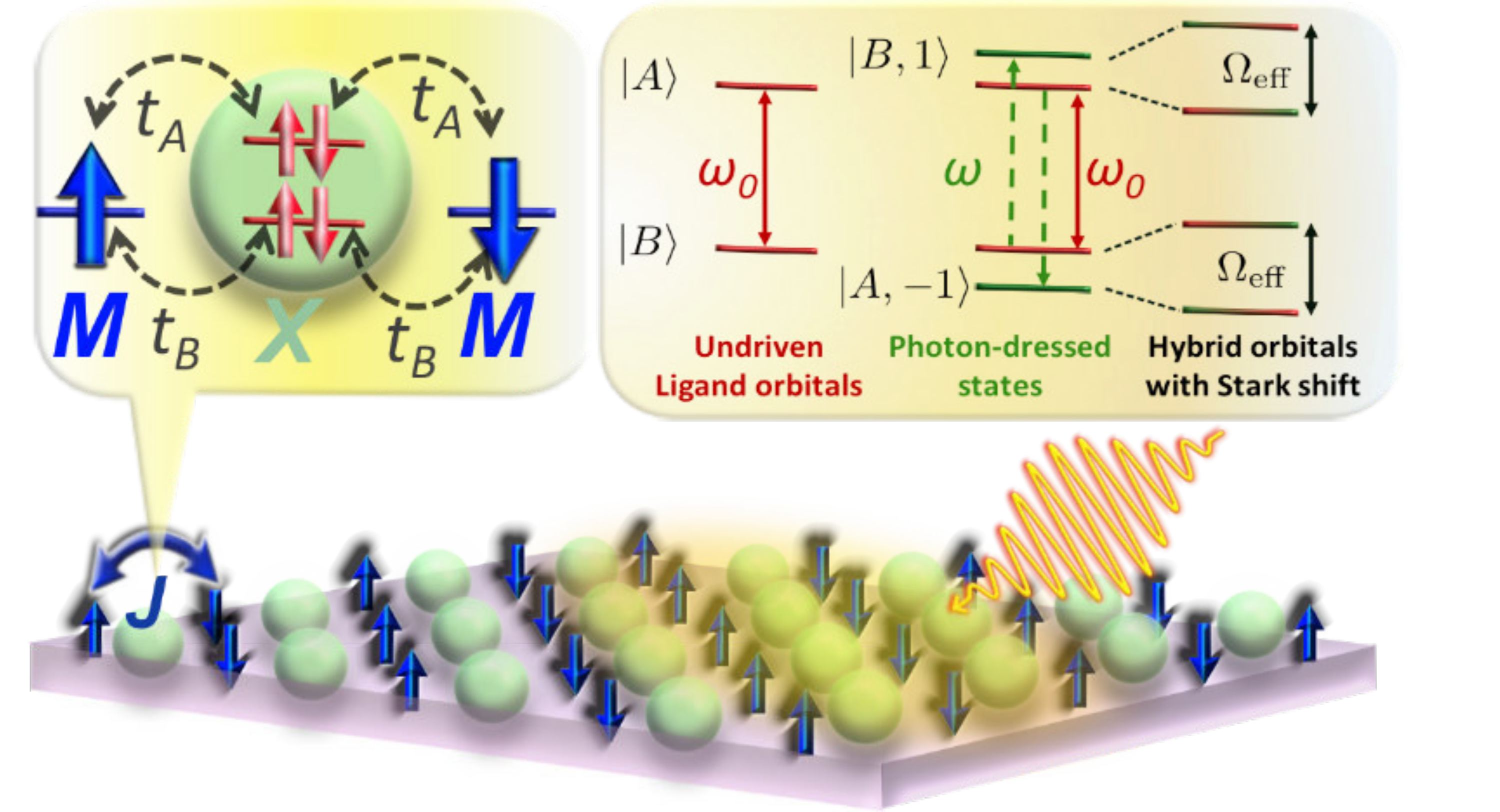}
	\caption{\label{driven}\textbf{Floquet engineering of spin exchange interactions using ligand orbitals: } Spin exchange interactions are typically mediated by non-magnetic ligand ions. Left Panel: Virtual hopping of electrons from one magnetic ion (M) to another via two orbitals (A and B) of the ligand ion (X). The magnetic coupling strength depends on the hopping parameter and the energy of the orbitals involved in this hopping process. Right Panel: In the presence of a periodic drive, these orbitals are replaced by hybridized photon-dressed orbitals (``Floquet replicas shown in green"). This splits the exchange channels and shifts the energies of virtual excitations, which modifies the exchange interactions.}
\end{figure}

\newpage
\textit{Floquet Engineering with ligand orbitals}.  In most magnetic materials, the spin-exchange interactions between two metal ions~(M) are mediated by  non-magnetic intermediary ligand ions~(X) as shown in Fig.~\ref{driven}. This superexchange occurs due to  virtual hopping of electrons within the cluster M-X-M. Therefore, the exchange interactions also depend on the properties of the orbitals of these non-magnetic ions. This dependence allows modifying the exchange interactions by manipulating the properties of the ligand orbitals involved in the exchange process. 

%The electronic energies of the orbitals involved in the exchange process, and  the hopping amplitude between the metal and the ligand ions affect the magnitude of the spin exchange interactions. Here, we explore the possibility of controlling the exchange interactions by manipulating the   properties of the ligand orbitals involved in the exchange process. 
%So far, in previous works on Floquet engineering of magnetic properties, the presence of the ligand ions is neglected, and direct hopping is assumed between the two metal sites. In these cases, the new exchange coupling depends on the renormalized hopping and the energy of the states which can be occupied during virtual excitations. Here, we explore the possibility of controlling the exchange interactions by manipulating the   properties of the ligand orbitals involved in the exchange process. 
%\begin{figure*}
%\centering
%\includegraphics[scale=0.4]{figs/orbital_virtual.pdf}
%\caption{The possible exchange pathways when the hopping between two metal sites is mediated via ligand orbitals. Gray panels show virtual intermediate states with their energies relative to the ground state with one spin on each metal site. Here, the ligand ion has two orbitals, and as a result there are many other channels available for spin exchange if the hopping between orbital B and metal sites is allowed.}% If $|E_B-E_A|\gg U$, and $t_A\approx t_B$, then the main contribution comes through $A$ orbitals only. }
%\label{virtual_excitation}
%\end{figure*}
The fact that a strong time-periodic drive which couples two orbitals of the ligand ion can modify the spin-exchange interactions follows from the Autler-Townes(AT) effect ~\cite{PhysRev.100.703}. In AT effect, a periodic drive results in the splitting of the absorbtion peak due to a change in the energy of the excited states. In our case,  the mixing of the two different ligand orbitals results in a change in the energy and the hybridization of the states available for the virtual excitations. These, in turn, alter the exchange interactions mediated by the ligand atoms. 
In order to study these changes, we consider a simple toy model with two metal ions with one spin on each, and a ligand ion with two filled orbitals which give rise to AF interaction between two spins at the metal ions. 
For the undriven case, the hamiltonian includes the hopping
between ligand orbitals (subscript $\alpha$) and metal sites (subscript $i$), an on-site spin interaction for each metal ion and the energy of the ligand
orbitals and is given by: %metal site spin interactions
\begin{equation}
H_{1}=H_{0}+H_{t}=\sum_{\alpha=A,B}\sum_{\sigma}E_{\alpha}n_{\alpha\sigma}+U\sum_{i}n_{i\up}n_{i\dn}+H_{t},
\end{equation}
and $H_{t}$ is the hopping between metal sites and intermediate orbitals given by:
\begin{equation}
H_{t}=-\sum_{i}\sum_{\alpha}t_{\alpha}c\dg_{\alpha\sigma}c_{i\sigma}+\text{{h.c}}
\end{equation}
with $|t_{\alpha}|\ll |E_{\alpha}|,U$. Assuming that each ligand orbital involved in the exchange process is completely filled, and on average there is one spin per metal site, the exchange energy can be calculated from  fourth-order perturbation theory by taking into account all the possible exchange pathways. Two such possibilities are shown in Fig. (1) of the Supplmental Material along with the energies of the virtual excitations. 
%When spin exchange interactions are mediated by a non-magnetic ion, the energies and shapes of the orbitals impact the hopping parameter and the energy of virtual excitations. This provides a control knob for the magnetic coupling strength between two metal ions.
The magnetic coupling strength~($J_{\text{ex}}$) upto fourth-order terms is given by:
\begin{equation}
\begin{split}
&J_{\text{ex}}=4\sum_{\alpha=A,B}t_\alpha^4\left(\frac{1}{\Delta_{\alpha }^2U}+\frac{1}{\Delta_{\alpha }^3}\right)+\frac{8t_A^2t_B^2}{\Delta_{A}\Delta_{B}U}\\&+4t_A^2t_B^2\left(\frac{1}{\Delta_{A }\Delta_{B }\Delta_{AB }}
+\frac{1}{\Delta_{A }^2\Delta_{AB}}+\frac{1}{\Delta_{B }^2\Delta_{AB }}\right),
\end{split}
\label{jex_two_orbital}
\end{equation}
where, $\Delta_{\alpha }=U-E_\alpha$ is the charge transfer gap and $\Delta_{AB}=(\Delta_{A }+\Delta_{B })/2$. In Mott insulators, $\Delta_{\alpha }\gg U$, and thus the exchange interactions reduces to:
\begin{equation}
J_{\text{ex}}\approx 4\frac{t_{\text{eff}}^2}{U},
\end{equation}
where, $t_{\text{eff}}=t_{\alpha}^2/\Delta_{\alpha }$ is the effective hopping between two magnetic ions induced by the ligand ion. 

Next, consider a periodic drive which can induce transitions between two ligand-ion orbitals:
\begin{equation}
H(t)=\Omega e^{-i\omega t}c\dg_{A\sigma}c_{B\sigma}+\Omega^{*}e^{i\omega t}c\dg_{B\sigma}c_{A\sigma}.
\label{orb_mix_drive}
\end{equation}
This kind of drive can be realized with an oscillating electric field $\textbf{E}(t)$, which couples orbitals $A$ and $B$ with strength $\Omega=\textbf{E}\cdot \textbf{P}/2$, where $\textbf{P}=e\left<A|\textbf{r}|B\right>$. 

 It modifies the orbitals involved in the spin exchange process as shown in Fig.~\ref{driven}, and as a result, not only do the energies of virtual excitations change, but they also increase in number, although the total weights sum up to the same value as the undriven case (see Supplemental Material). % \color{blue}{Gil: although not in total weight}
%\\ \color{red}{Yes, total weights sum up to the same value as the undriven case, but the energies of virtual excitations are different which result in a different spin exchange coupling.}
%\color{black}
%I THINK IT CAN BE MADE BETTER BY STATING - FLOQUET SPACE HYBRIDIZATION AND STRAK EFFECT
In the presence of a periodic drive, the complete Hamiltonian, $H=H_{0}+H_{t}+H(t)$, can now be treated using an extended Floquet basis, i.e, the direct product of the states in the actual Hilbert space and the photon number states. We treat the hopping part, $H_t$, as a perturbation similar to as in the undriven case, but now the virtual excitations are the energy eigenstates of the full Floquet hamiltonian describing $H_{0}+H(t)$. We choose the drive parameters such that the effective-spin hamiltonian picture remains valid. The periodic drive in Eq.~(\ref{orb_mix_drive}) mixes the ligand orbitals $A$ and $B$, and the virtual excitations now involve the hybrid Floquet orbitals given by: 
%subspace of eigenstates with single spin on each metal site is well separated in energy from the rest of the eigenstates.  
\begin{equation}
\begin{split}
&\Ket{P,n}=\cos\frac{\theta}{2}\Ket{A,n}+\sin\frac{\theta}{2}\Ket{B,n+1},\\ &\Ket{M,n}=\sin\frac{\theta}{2}\Ket{A,n}-\cos\frac{\theta}{2}\Ket{B,n+1},
\label{PM}
\end{split}
\end{equation}
where $\cos\theta=\frac{\delta}{\sqrt{\delta^{2}+4\Omega^{2}}},\,\sin\theta=-\frac{2\Omega}{\sqrt{\delta^{2}+4\Omega^{2}}}$, $\delta=\omega-\omega_0$ is the detuning, $n$ denotes the photon index, and~$\omega_0=E_B$-$E_A$ is the energy difference between two ligand orbitals. %OMEGA_0

Once again, the magnetic-coupling strength can be calculated using fourth-order perturbation theory or by diagonalizing the Floquet hamiltonian numerically. The expression for the new magnetic coupling $J_{\text ex}$ is similar to that in Eq.~(\ref{jex_two_orbital}), with orbitals $A$ and $B$ replaced by their hybrid counterparts $\ket{P,n}$ and $\ket{M,n}$ (see Supplemental Material for the derivation). Since, the hopping parameter and the energy of these orbitals are different from the undriven case, the exchange interactions are modified. 

\begin{figure}[H]
	\centering
\includegraphics[scale=0.5]{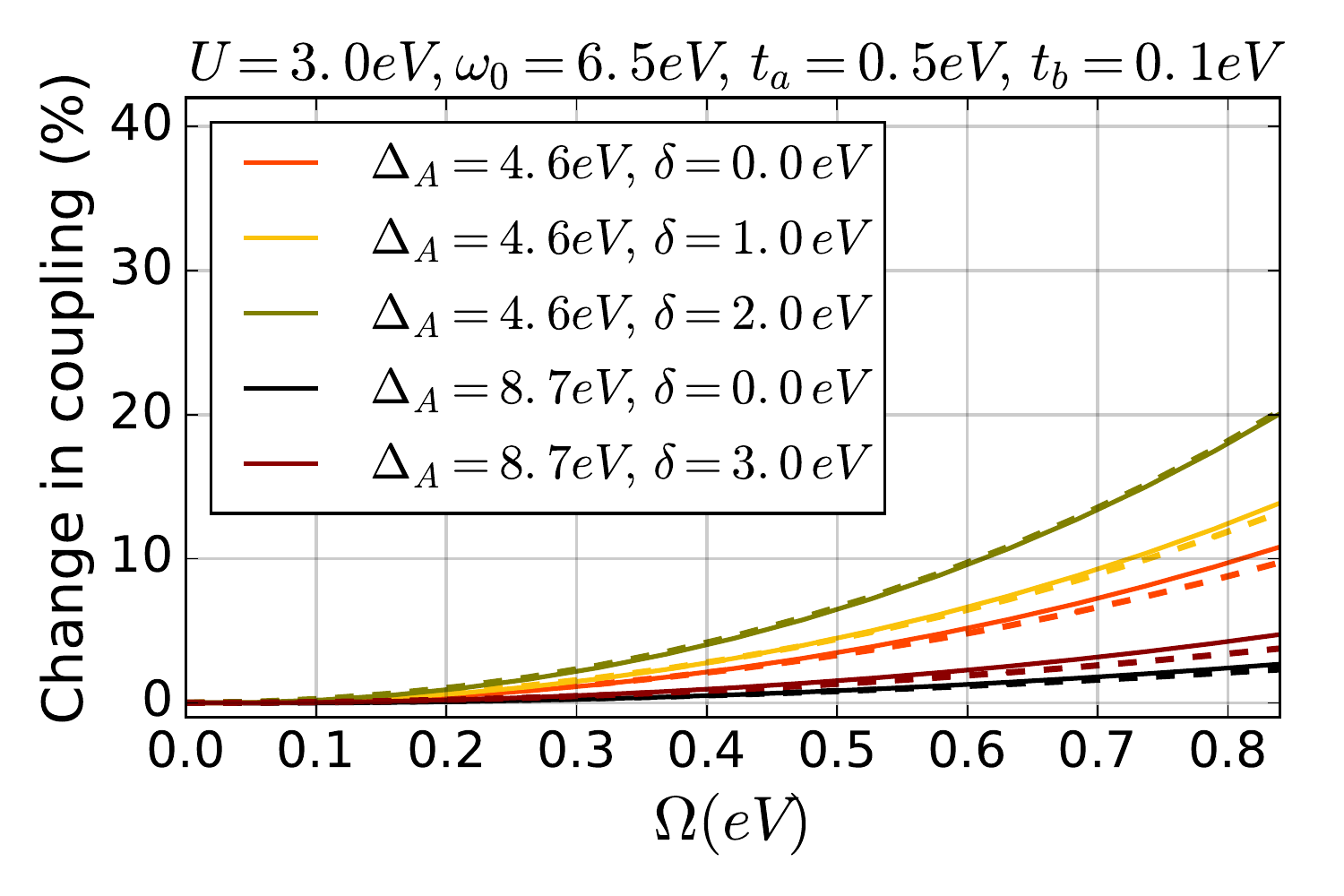}
\caption{Change in magnetic coupling as a function of drive strength $\Omega$ from numerics(solid lines) and theory(dashed lines) where the periodic drive mixes two orbitals of the ligand ion. The effect of the drive is large when the effective Rabi frequency is comparable to the charge transfer gap $\Delta_{A}$. These parameters were chosen according to the typical values of interaction energy $U$ and hopping parameter for TMTCs. }
\label{hybridorbital}
\end{figure}

These changes are shown in Fig.~\ref{hybridorbital}, where we also plot the results obtained from the numerical diagonalization of the Floquet hamiltonian with four Floquet zones included. Clearly, the exchange interactions can be modified significantly by tuning the drive parameters. This effect arises due to the changes in energy and number of virtual excitations which increase due to the splitting of channels available for spin-exchange processes. This splitting occurs as virtual excitations now belong to two different Floquet sectors shown in right panel of Fig.~\ref{driven}. This change in coupling strength is significant only if the Rabi splitting $\Omega_{\text{eff}}$, between two states in each Floquet sector is of the same order as the charge transfer gap~$\Delta_{A i}$. 

%U=3,delta=6.7
%few more drafts.
%\textbf{toy model - only for two metal sites, RWA assumed}\\
%\textbf{long expression -put it in appendix}
%\textbf{If we add the possibility of photon-assisted tunneling, we can access other Floquet zones and that would open up many more channels for spin exchange - so it is worth exploring.}

%drive frequency 

%\subsubsection{Suitable Regime}
As shown in~Fig.~\ref{hybridorbital}, the modifications in the exchange interactions are significant only if the shift in energy levels of the states available for virtual excitations is comparable to the charge-transfer gap.  Usually, the charge transfer gap $\Delta_{A i}\approx 5$-$10$eV, and thus in order to observe any significant effect, we need $\Omega\approx 1$eV. One of the most common ligands is Sulphur ion where one can consider mixing $3s$ and $3p$ orbitals where the energy difference between two orbitals, $\omega_0\approx5$-$10$eV, and the dipole moment matrix element $|\textbf{P}|\approx0.6e\AA$ (see Supplemental Material), and thus we need $E\approx5\text{V}/\AA$ to get a $\Delta J/J\approx10\%$. The materials with small charge transfer gap, large dipole moment matrix element, and small energy gap ($\omega_0$) are ideal candidates for this scheme to work at lower electric field.

% {\bf This effect is really for direct exchange modification - different than above. This should be said somewhere}
% \todo[inline]{I found the usage of this term  quite confusing.  I noticed that the direct exchange term is used mostly for  Hund's coupling like interactions. Usually, the e-e coulombic interactions give rise to two kind of terms - 1. on-site coulombic interaction and 2. intra-atomic or interatomic  exchange interactions (Hund's like). In the present case, although there is no ligand ion, these interactions arise mainly because of hopping (kinetic energy terms) and hence I prefer to understand it as superexchange from direct hopping. I found this notation very confusing, and many people use the term superexchange only for ligand mediated magnetic interactions but this should also include the direct hopping.}
\textit{Floquet Engineering with metal ion orbitals}. In  the toy model above we assumed only a single orbital for each TM ion. The magnetic properties of TM compounds, however, are significantly affected by the occupancy of other $d$ orbitals, crystal field splitting, and on-site interactions within these $d$ orbitals. %Since, the energy of these virtual excitations depends on the energies and the overlap of orbitals from two different sites.
We can change the properties of these orbitals by using a periodic drive. This kind of drive would result in an AC Stark shift of the  energy levels in both singly and doubly occupied sectors. As a result, the virtual excitations would now involve the hybrid orbitals, and hence the magnetic coupling strength would change (see Supplemental Material for more details). This modification occurs only if the Stark shift is different for the low energy subspace and the states available for virtual excitations. 
%{\bf Is it really Rabi splitting? Maybe Rabi splitting of the resonances?}  

%t is mainly because if the AC stark shift is same for the ground state and the virtually excited state, then the energy of virtual excitations is not affected. 

We study the effect of the orbital mixing with a simple toy model where magnetic interactions arise from direct hopping between two TM ions. We consider a two-site Fermi-Hubbard model with two orbitals on each site and at quarter filling, in the presence of a periodic drive which couples the two levels on each site.  It can be represented by the following hamiltonian:
\begin{equation}
H=H_t+H_k+H_0,
\label{fullH_metalsite}
\end{equation}
where $H_t$ is hopping term given by:
\begin{equation}
\begin{split}
H_t=-\sum_{\sigma,\alpha=A,B}t_\alpha c_{1\alpha\sigma}\dg c_{2\alpha\sigma}-t_{AB}\sum_{\sigma,i\ne j}c_{1A\sigma}\dg c_{2B\sigma}+\text{h.c},
\label{metal_site_hopping}
\end{split}
\end{equation}
$H_k$ is the on-site Kanamori interaction~\cite{kanamori1963j}:
\begin{equation}\begin{split}
H_k=U\sum_{i,\alpha}&\hat{n}_{i\alpha\uparrow}\hat{n}_{i\alpha\downarrow}+U_1\sum_{i,\alpha<\beta,\sigma,\sigma'}\hat{n}_{i\alpha\sigma}\hat{n}_{i\beta\sigma}\\&-J_H\sum_{i,\alpha<\beta,\sigma,\sigma'}c_{i\alpha\sigma}\dg c_{i\alpha\sigma'}c_{i\beta\sigma'}\dg c_{i\beta\sigma},
\end{split}
\end{equation}
and the on-site  energy
\begin{equation}
H_0=\sum_i E_A(\hat{n}_{iA}-1)+(E_A+\omega_0)\hat{n}_{iB}
\end{equation}
with $U,\,U_1\gg t_\alpha$. At quarter filling, if $\omega_0\gg \frac{t_\alpha^2}{U}$, then the low-energy subspace consists of states with one spin in each $A$ orbital, and the magnetic coupling strength is approximately given by $J_{ex}=4t_A^2/U$. On the other hand, if $\omega_0=0$ and $t_{ab}=0$, the ground state is FM in spin but AF in the orbital degree of freedom. Here, we are mainly interested in the first scenario, which allows us to mix two orbitals by applying a periodic drive of the form:
\begin{equation}
H(t)=\sum_{i,\sigma}(\Omega e^{i\omega t}c_{iA\sigma}\dg c_{iB\sigma}+\Omega^*e^{-i\omega t}c_{iB\sigma}\dg c_{iA\sigma}).
\label{metalsite_mixing}
\end{equation}
\begin{figure}[H]
	\includegraphics[scale=0.24]{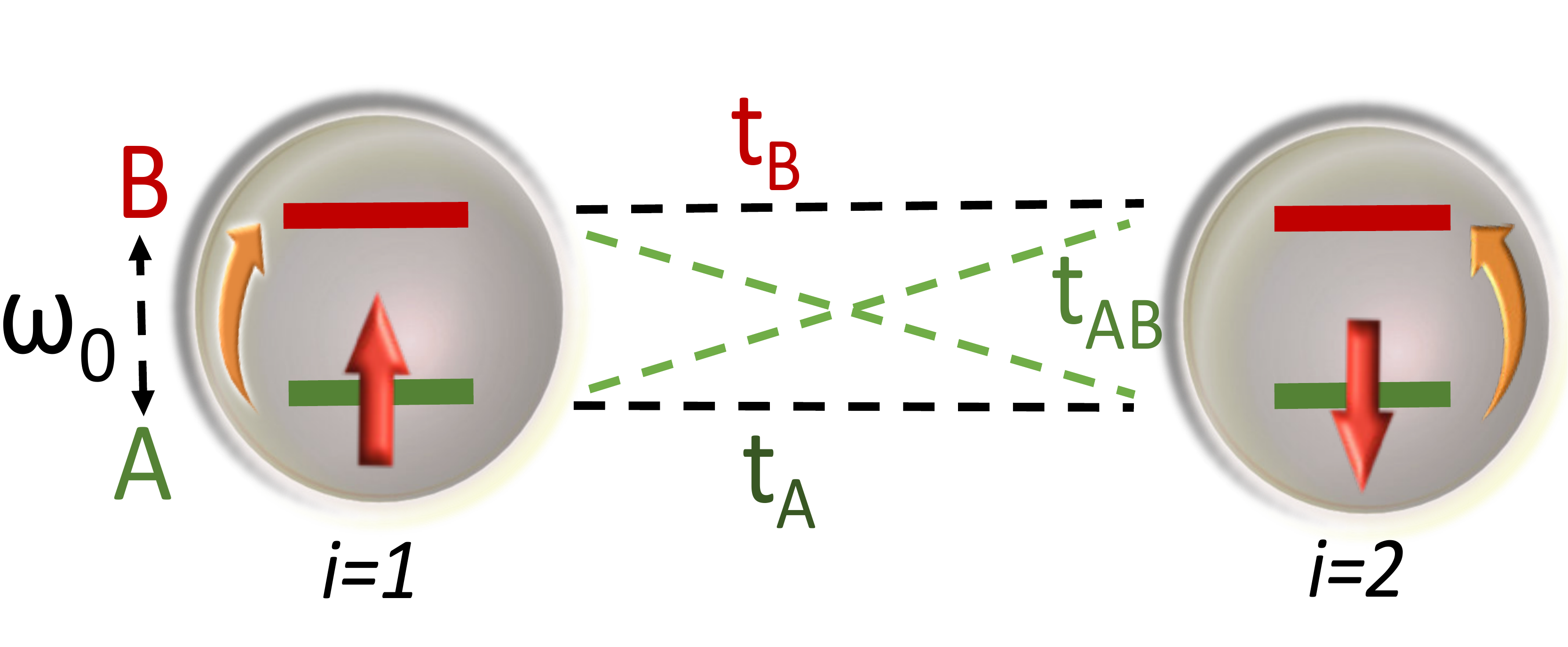}
	\caption{\textbf{Schematic for metal orbital Floquet Engineering : }A two-site Fermi-Hubbard model with two orbitals on each site at quarter filling. Two orbitals denoted by A and B with electronic energy $E_A$, and $E_A+\omega_0$ are mixed using a periodic drive given in Eq.~(\ref{metalsite_mixing}). For simplicity, we assume direct hopping between two metal ions.}
\end{figure}
Let us now study the changes in the Floquet eigenstates connected to the low lying energy subspace  of the undriven hamiltonian as a function of different drive parameters. We focus mainly on the regime where the effective spin picture is valid and calculate the spin exchange interactions from the energy difference between singlet and triplet states (details in Supplemental Material). As shown in Fig.~\ref{compare}, the magnetic coupling strength can change significantly depending on the frequency and strength of the drive. 

%This scheme can be realized in those TM compounds where exchange processes involve only one electron from each TM ion, and there are some empty $d$ orbitals available for mixing. 
This scheme can be realized in those magnetic materials where TM ions have $d^1$ configuration. In transition metal compounds with octahedral or tetrahedral ligand cages, $d$ orbitals  split into $e_g$ and $t_{2g}$ levels with crystal-field splitting parameter in the range of $0.3\,eV$  to $1.5\,eV$. The periodic drive can be realized with an AC electric field which couples these \textit{d} orbitals. Therefore, the drive amplitude is $\Omega=e\left<\psi_A|\textbf{E}\cdot\textbf{r}|\psi_B\right>/2$.
\begin{figure}[H]
	\centering
	
	\includegraphics[scale=0.32]{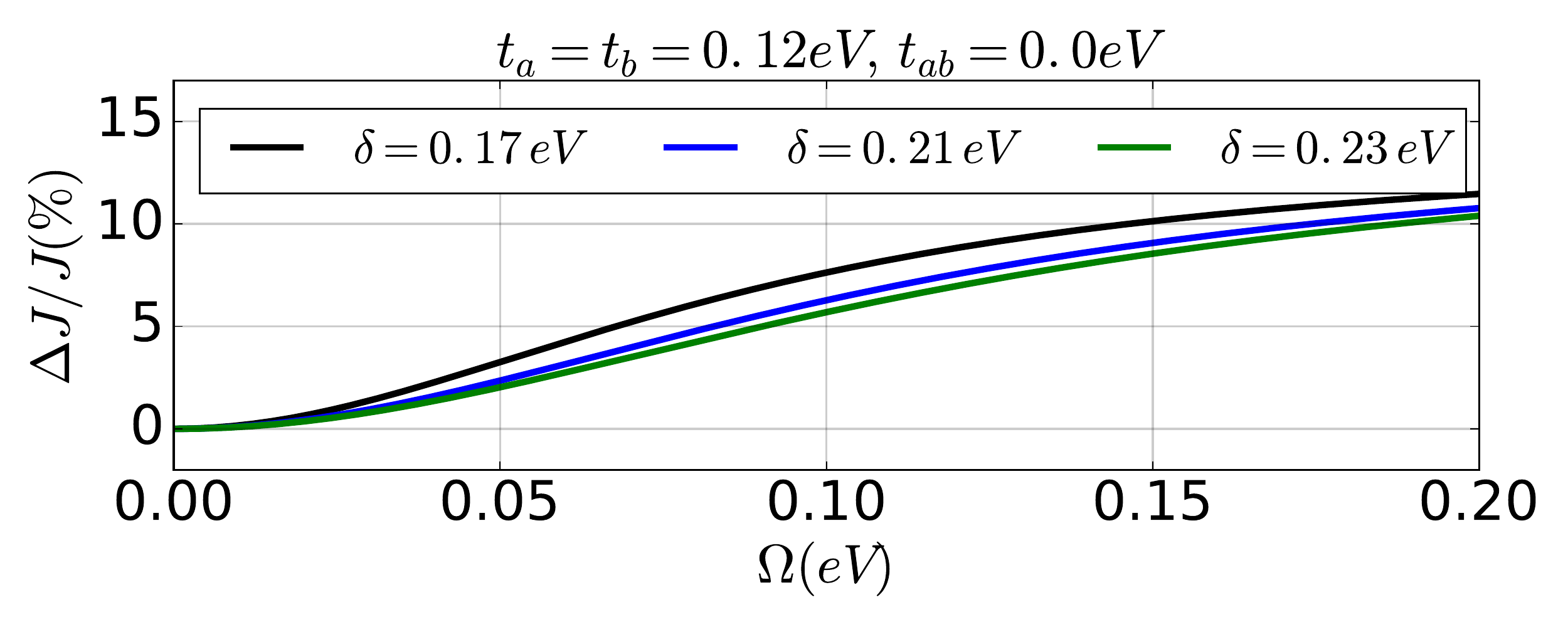}
	\includegraphics[scale=0.325]{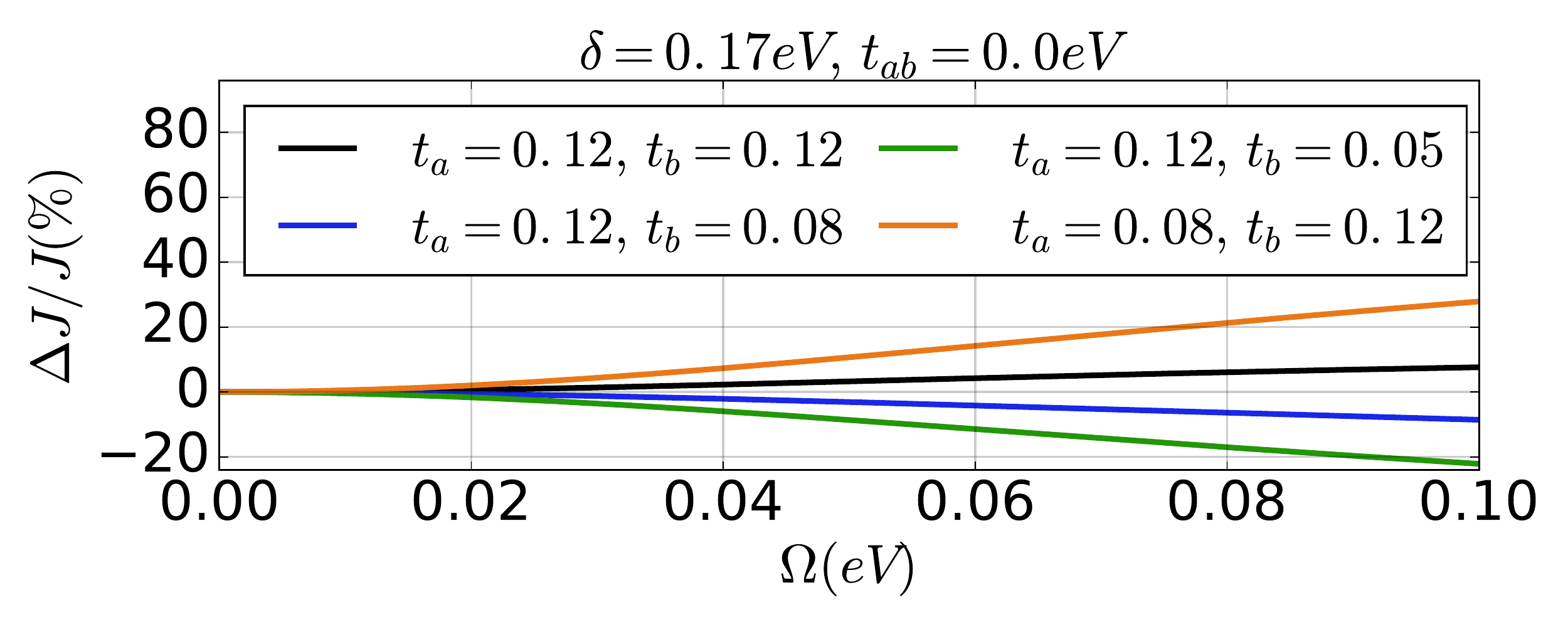}
	\includegraphics[scale=0.32]{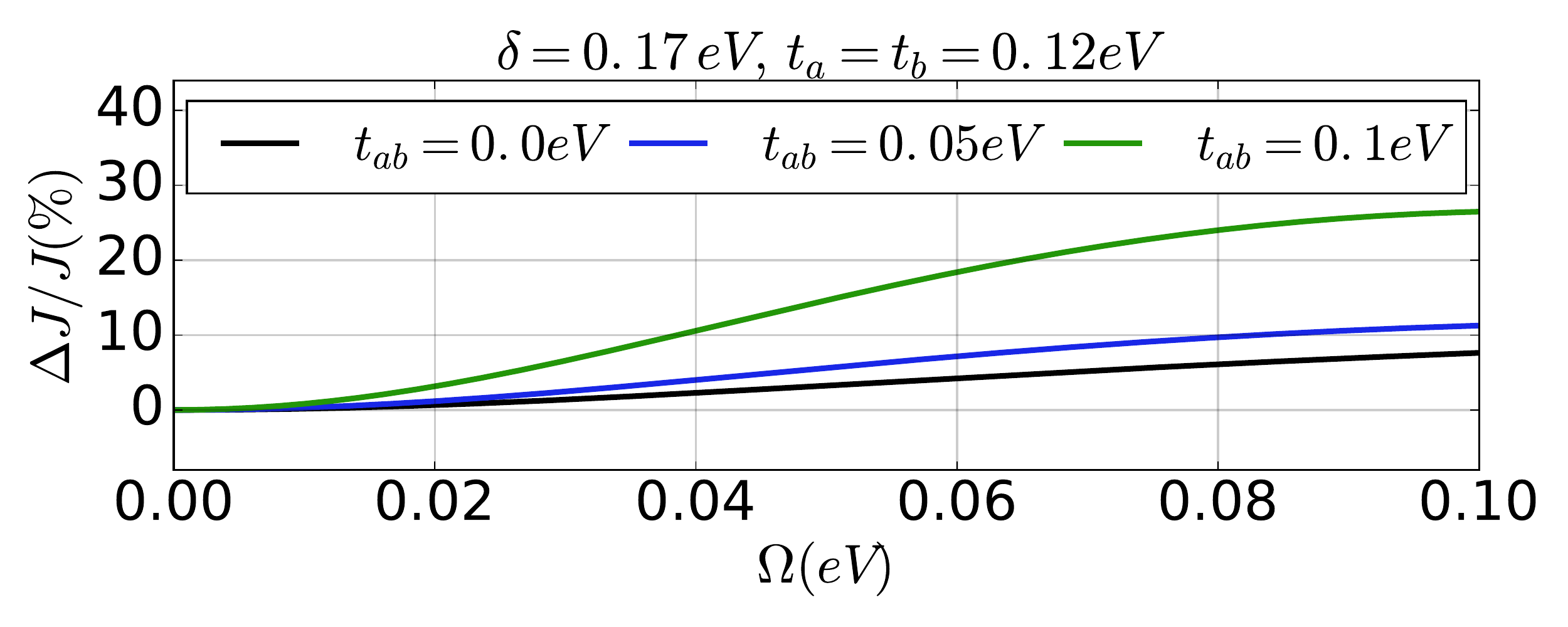}
	\includegraphics[scale=0.32]{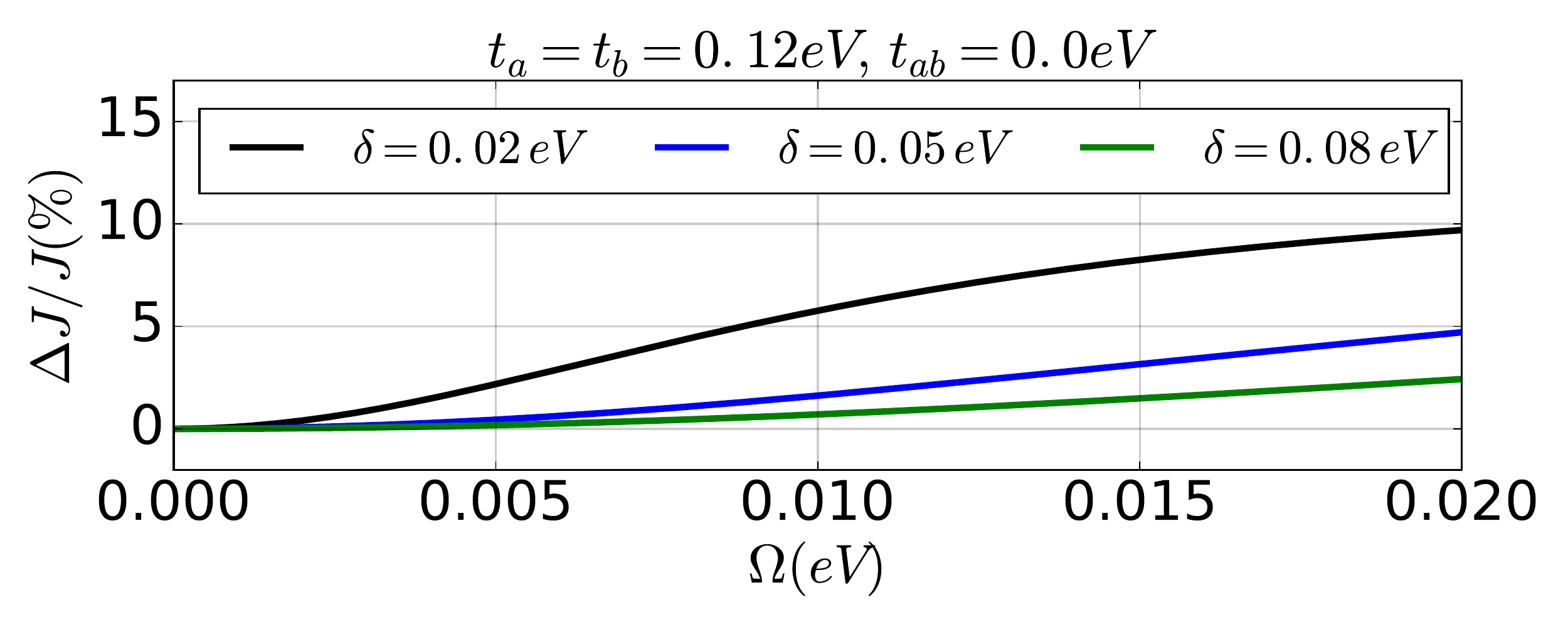}
	\caption{	\label{compare}Effect of different parameters on the change in magnetic coupling strength as a function of drive amplitude $\Omega$ for $U=4.0eV$, $J_H=0.8eV$, $U_1=U-2J_H$, and $\omega_0=0.91eV$. These changes are large when the detuning is decreased. The second panel shows that large imbalance between $t_a$ and $t_b$ makes these changes more prominent. Similarly, we also observe that large $t_{ab}$ results in large changes. In the last panel, we show the changes for a very small detuning where a significant change can be seen at extremely small drive amplitudes.}
	% \textbf{Large Omega - in terms of eV/Ang}
\end{figure}
 The only orbitals involved in this transition, however, are $d$ orbitals, and the dipole transitions between same-parity orbitals are forbidden. Nevertheless, the crystal field can give rise to \textit{d-p} mixing in non-centrosymmetric compounds which also renders some $p$ character to the otherwise pure $d$ orbitals, and thus such dipole transitions are now weakly allowed. For some tetrahedral complexes, this mixing is $1$-$5\%$~\cite{dpballhausen1958intensities}, and thus $d$-$d$ dipole moment matrix element $|\left<d_i|\textbf{r}|d_j\right>|\approx0.05\text{e\AA}$ which corresponds to a drive strength $\Omega\approx0.02eV$ at $E=1V/\AA$.  Although, there are some magnetic materials where some of the TM ions are surrounded by tetrahedral cage of ligand ions~\cite{PhysRev.102.1008}, at this stage, we are not aware of any such magnetic materials where the TM ion with $d^1$ configuration is surrounded by a tetrahedral arrangement of ligand ions.
 On the other hand, in octahedral geometry, some mechanisms like coupling with vibrational modes, and mixing with ligand \textit{p} orbitals~\cite{dpnaiman1961interpretation, dpliehr1957intensities,dpballhausen1958intensities,dpballhausen1959intensities} allow these \textit{d-d} transitions.  This d-p mixing  can be estimated from the oscillator strength of $d$-$d$ transitions in octahedral complexes (Table I of Ref.\cite{dpliehr1957intensities}), and it is roughly of the order of $0.1\%$. This corresponds to a \textit{d-d} electric dipole moment matrix element, $P=e\left|\left<d_{t_{2g}}\right|\textbf{r}\left|d_{t_{e_g}}\right>\right|\approx0.01\text{e\AA}$, and thus the drive strength, $\Omega\approx0.005\text{eV}$ for $E=1V/\AA$. It indicates that a change of $10\%$ in magnetic coupling can be achieved only at $E\approx\text{5-10} \text{eV/\AA}$. 
 
 %%%%%%This offers an advantage for tertrahedral complexes....element check once again
 %This scheme can be applied to some Mott insulating rare-earth titanates ($\text{RTiO}_3$), where the degeneracy between  different $t_{2g}$ orbitals is lifted~\cite{solovyev2006lattice} due to some octahedral distortions. 
 
 %There is not a very clear consensus on this issue.
%This kind of transitions although very weak, but  have been observed in transition metal trichalocgenide $\text{NiPS}_3$~\cite{dpPhysRevLett.120.136402}, which indicate the possibility of using this scheme in TMTCs
%\subsubsection{Other routes to orbital hybridization}

%\todo[inline]{In octahedral materials d-d dipole moment is extremely small, and now I have added a line above. This is because the d-p mixing factor ~0.01 which I calculated earlier using vibronic coupling mechanism. }

%
The performance of the metal orbital hybridization scheme depends on our ability to mix the two $d$ orbitals with light. In addition to the dipole transition,  this kind of mixing can also be achieved by employing two-photon processes or by direct vibrational coupling between two levels. For a two-photon process between two $3d$ orbitals, the drive amplitude depends on the dipole moment between $d$ and the other odd parity orbitals, and it is proportional to the intensity of the EM field.  For such processes, the matrix element between two $d$ orbitals is given by $\Omega\approx e^2 E^2P_{dd}$, where $P_{dd}\approx\frac{1}{2}\frac{|\left<3d|\textbf{r}|4p\right>/2|^2}{E_{4p}-E_{3d}}\approx10^{-3}\AA^2/\text{eV}$, and thus $\Omega\approx 10^{-3}\text{eV}$ for electric field, $E\approx 1\text{eV/\AA}$. Similarly, one can also make use of coherent lattice vibrations to achieve a similar hybridization between two $d$ orbitals. In perfect octahedral symmetry, the direct vibrational coupling between  some $d$ orbitals can occur  for those Raman active modes which involve metal-ligand bond rotation, i.e, $T_{2g}$ phonon modes.
%\cite{PhysRevB.98.024102}
%In perfect octahedral symmetry, the direct vibrational coupling between  $e_g$ and $t_{2g}$ orbitals can occur only for those Raman active modes where ligand ions move perpendicular to the metal-ligand bonds, i.e $T_{2g}$ phonon modes.
 Usually, this kind of motion is associated with phonons in the frequency range of 50-100meV, and thus it might be applied to materials where the energy difference between two $d$ orbitals is in the same range. This scheme can be used in some rare-earth titanates ($\text{RTiO}_3$), where even $t_{2g}$ bands are non-degenerate with a crystal-field splitting $\Delta_{\text{CF}}\approx \text{30-400} \text{meV}$~\cite{solovyev2006lattice}, and some phonon modes (e.g. $A_g(2),A_g(4),B_{1g}(3),B_{1g}(4),B_{2g}(4) $) which involve the bond rotations have frequencies ranging from $\text{10-100}\text{meV}$~\cite{PhysRevB.57.2872,PhysRevB.69.172301}. %reedyk1997raman}. 
In this scheme, the drive strength depends on the phonon amplitude, and $\Omega\approx0.03\text{eV}$ (Sec. V of Supplemental Material) for a lattice displacement of $0.1\text{\AA}$. %(phonon amplitude ~0.6AngSqrt[amu]-typical experimental values - 2)
This kind of phonon motion is possible in $\text{LaTiO}_3$ by making use of large nonlinear phononic interactions of these modes with some infrared phonon modes where a lattice displacement of $0.1\text{\AA}$ can be achieved at electric field, $E\approx0.2V/\AA$~\cite{ PhysRevB.89.220301,PhysRevB.98.024102}. This corresponds to a change of magnetic coupling by $\text{5-10}\%$.

%These schemes work in very different energy regimes, and we compare different methods in Fig.~\ref{schemecomparison} for different values of the drive parameters assuming a small energy gap ($\omega_0\approx5 \text{eV}$) and dipole moment $P=0.5\text{e\AA}$ for the ligand orbitals, and an energy gap, $\omega_0\approx1 \text{eV}$ for the metal orbital mixing with dipole moment $P\approx0.1\text{e\AA}$. Although, the principles behind the photo-modified direct hopping and orbital mixing are very different, significant changes are observed at E fields of the same order of magnitude. 
%rough estimate for phonon in metal orbital mixing scheme - fluence~5mJ/cm^2, pulse duration ~100 fs, Q~0.5AngSqrt(amu)~0.05Ang, and E~3MV/cm. furthermore in the linear regime, the lattice displacement is directly propertional to the fluence-a displacement of 0.5Ang-300MV/cm.

\textit{Conclusions}. To summarize, we provide a novel protocol to control the magnetic properties of materials by manipulating the orbital degrees of freedom with light.  In the previous  works~\cite{Mentink2014,mentink2015ultrafast,mentink2017manipulating,Bukov2016,hejazi1PhysRevLett.121.107201,hejazi2018floquet}, spin-exchange interactions change due to photo-assisted hopping while in our case, similar effects originated due to AC Stark shift of the levels available for virtual excitations. The ligand orbital mixing scheme and the photo-modified direct hopping gives significant changes only at $E\approx\text{1-5}\text{eV/\AA}$.
In the case of  photo-modified direct hopping, changes are very sensitive to the values of the drive parameters, but for the orbital mixing case,  spin-exchange interactions change monotonically with drive parameters as long as the perturbation method is valid. For the ligand orbital mixing scheme, the typical energy gap between $s$-$p$ orbitals of the ligand ions is 5-10eV, and thus it works at high frequency, but the photo-modified direct hopping works well at frequencies in the range $\text{0.5-2}\text{eV}$. This ligand orbital scheme thus broadens the frequency range for the applicability of Floquet engineering of spin-exchange interactions. On the other hand, the metal orbital scheme involving a phonon drive should give a change of $5$-$10\%$ at much smaller frequencies, and $E\approx\text{0.2-0.5}\text{V/\AA}$ which is about ten times smaller than the electric field required for other schemes. We showed that controlling the orbital degrees of freedom with light opens up new possibilities for coherent manipulation of properties of quantum materials.

\textit{Acknowledgements}. We acknowledge support from the IQIM, an NSF physics frontier center funded by Gordon and Betty Moore foundation. We are grateful for support from ARO MURI W911NF-16-1-0361 ``Quantum Materials by Design with Electromagnetic Excitation" sponsored by the U.S. Army.

\bibliographystyle{apsrev4-1}   %use this part if you are using revtex4-1
\bibliography{magnetismref.bib}

\newpage

%\section{Introduction
\onecolumngrid
\section*{Supplemental Material for ``Orbital Floquet engineering of Exchange Interactions in Magnetic Materials"}
\twocolumngrid
\section{Review : Toy Model for AFM coupling renormalization due to photo-modified direct hopping }
%prototype Mott-Hubbards
We briefly review the effect of a periodic drive on the exchange interactions using the periodically driven Fermi Hubbard model (FHM) in Mott regime at half-filling.
In the presence of a time dependent electric field, the full Hamiltonian of the Fermi-Hubbard model is given by:
\begin{equation}
	H=-t\sm c_{i\sigma}\dg c_{j\sigma}+\text{h.c}+U\sum_{i}n_{i\uparrow}n_{i\downarrow}+\mathbf{E}\cdot\mathbf{\sum_{i,\sigma}}n_{i\sigma}\mathbf{r_{j}\cos(\omega t)}.
\end{equation}
After Peierls substitution, it becomes:
\begin{equation}
	H'=-t\sm e^{i\left[\mathbf{\frac{\mathbf{E\cdot}(\mathbf{r_{j}-}\mathbf{r}_{i})}{\omega}}\sin(\omega t)\right]}c_{i\sigma}\dg c_{j\sigma}+\text{h.c}+H_{U}=H_{t}'+H_{U}.
\end{equation}
In the limit $U\gg t$, and for a non-resonant drive, the exchange coupling is given by:
%\begin{equation}
%J_{i}'=J_{i}U\left(\frac{1}{U}J_{0}(\xi_{i})^{2}+\frac{1}{U+\hbar\omega}J_{1}(\xi_{i})^{2}+\frac{1}{U-\h%bar\omega}J_{1}(\xi_{i})^{2}+\frac{1}{U+2\hbar\omega}J_{2}(\xi_{i})^{2}+\frac{1}{U-2\hbar\omega}J%_{2}(\xi_{i})^{2}+.......\right)
%\label{_hoppingrenormalization}
%\end{equation}
\begin{equation}
	J_{i}'=J_{i}U\sum_{n=-\infty}^{\infty} \frac{1}{U+n\omega}\mathcal{J}_n(\zeta_i)^2,
	\label{hoppingrenormalization}
\end{equation}
where, $J_i=\frac{4t^2}{U}$ is the magnetic coupling strength for the undriven case,  $\mathcal{J}_n$ denotes $n^{th}$ order Bessel function, and drive parameter
\begin{equation}
	\zeta_i=\frac{\textbf{E}\cdot(\textbf{r}_j-\textbf{r}_i)}{\omega}.
\end{equation}
In the presence of this periodic drive, the spin exchange interactions are affected mainly due to two factors: (a) change in the hopping parameter due to photon-assisted tunneling  (b) virtual excitations between different Floquet sectors as shown in Fig.~3 of Ref.~\cite{mentink2017manipulating}. As a result, the effective spin exchange interactions can be controlled by changing the frequency, polarization and intensity of the laser. Previous works~\cite{Mentink2014,mentink2015ultrafast,mentink2017manipulating,Bukov2016,hejazi1PhysRevLett.121.107201,hejazi2018floquet} have studied the periodically driven FHM extensively for both resonant and off-resonant cases. The above expression in Eq.~(\ref{hoppingrenormalization}) is valid only for a non-resonant drive where doublon sectors are well separated in energy from the single occupation sector. Resonant drive can be handled using a somewhat similar machinery of Floquet formalism as shown in Ref.~\cite{Bukov2016}. For a near resonant drive, real doublon-holon pairs can significantly affect the exchange interactions and its effects are discussed in great details in Ref.~\cite{Liu2018}. 
\section{AFM exchange via two orbitals of the same ligand ion in the presence of a periodic drive}
\label{two_orbital_mixing_ligand_appendix}
%expression for undriven case
%expression for orbital mixing
%how it changes once you also take into account the photon-assisted hopping.

When the hopping between two metal sites is allowed via two orbitals of the ligand ion,  spin exchange energy for undriven case is given by :
\begin{equation}
	\begin{split}
		&J_{ex}=4\sum_{\alpha=A,B}t_\alpha^4\left(\frac{1}{\Delta_{\alpha i}^2U}+\frac{1}{\Delta_{\alpha i}^3}\right)+\frac{8t_A^2t_B^2}{\Delta_{Ai}\Delta_{Bi}U}\\&+4t_A^2t_B^2\left(\frac{1}{\Delta_{A i}\Delta_{B i}\Delta_{AB i}}
		+\frac{1}{\Delta_{A i}^2\Delta_{ABi}}+\frac{1}{\Delta_{B i}^2\Delta_{AB i}}\right),
	\end{split}
	\label{jex_two_orbital}
\end{equation}
where, $\Delta_{\alpha i}=U-E_\alpha$ is the charge transfer gap, and $\Delta_{AB i}=(\Delta_{A i}+\Delta_{B i})/2$. This expression was obtained by applying fourth order perturbation theory to the following hamiltonian:
\begin{equation}
	\begin{split}
		H_{1}=H_{0}+H_{t}=&\sum_{\alpha=A,B}\sum_{\sigma}E_{\alpha}n_{\alpha\sigma}+U\sum_{i=1,2}n_{i\up}n_{i\dn}\\&-\sum_{i=1,2}\sum_{\alpha}t_{\alpha}c\dg_{\alpha\sigma}c_{i\sigma}+\text{{h.c}}
	\end{split}
\end{equation}
where $\alpha=A,B$ are two orbitals of the ligand ion involved in the process of superexchange between the spins at two metal sites denoted by $i=1,2$ above, and the hopping parameter $t_{A/B}\ll U,|E_\alpha|$. 
\begin{figure*}
	\centering
	\includegraphics[scale=0.5]{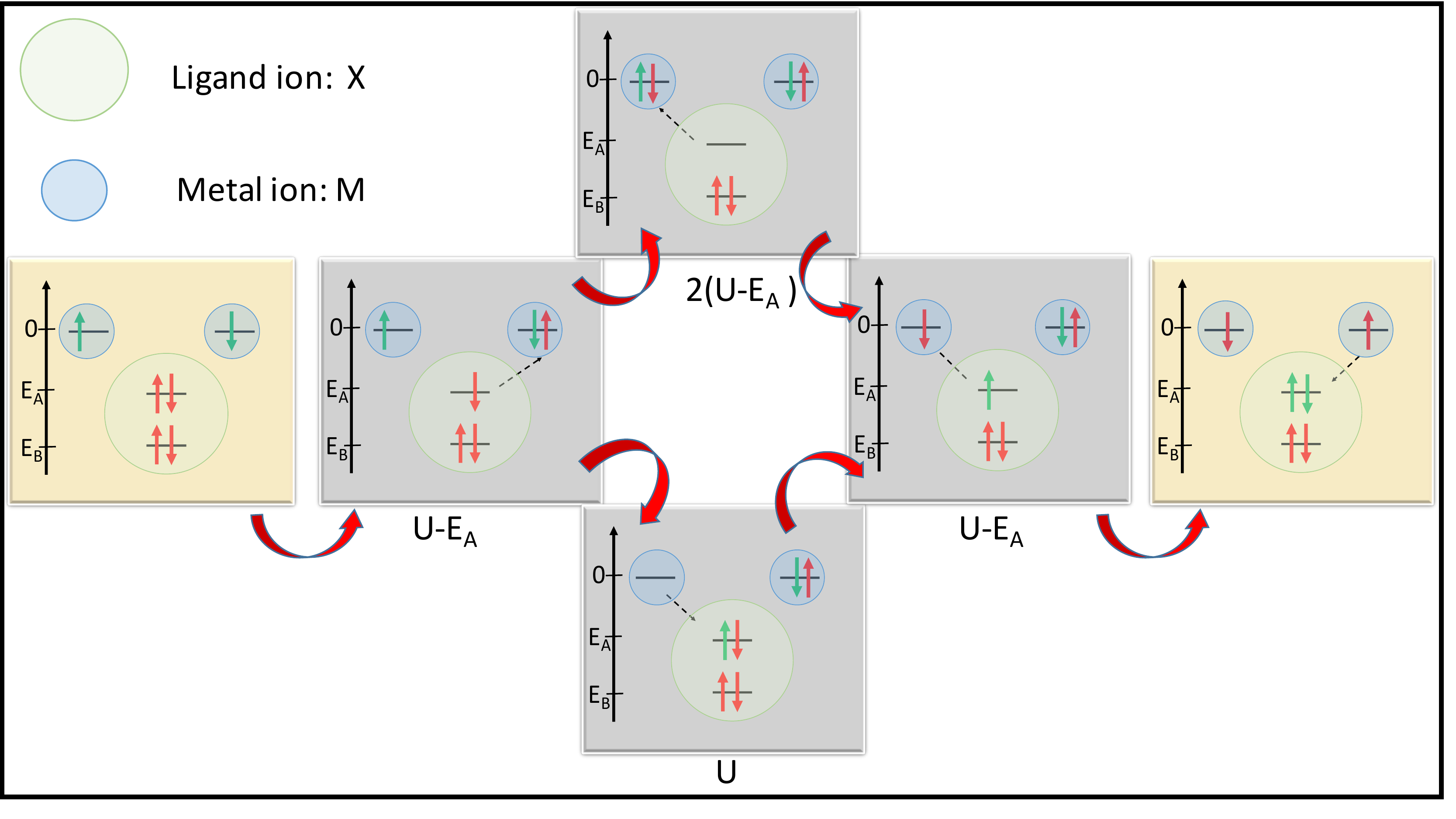}
	\caption{Two possible spin exchange processes when the hopping between two metal sites is mediated via ligand orbitals. Gray panels show virtual intermediate states with their energies relative to the ground state with one spin on each metal site. Here, the ligand ion has two orbitals, and as a result there are many other channels available for spin exchange if the hopping between orbital B and metal sites is allowed.}% If $|E_B-E_A|\gg U$, and $t_A\approx t_B$, then the main contribution comes through $A$ orbitals only. }
	\label{virtual_excitation}
\end{figure*}
%\begin{figure}

%	\includegraphics[scale=0.3]{figs/ligand_ion_mixing2.pdf}
%	\caption{This figures shows two metal sites at $i=1,2$ with one orbital, and one spin at each site, and a ligand ion whose two orbitals A and B are involved in the superexchange process. These two orbitals have different electronic energy, and a periodic drive (Eq. (\ref{orb_mix_drive})) mixes these two orbitals and thus alter the spin exchange interactions.  }
%	\label{ligand_mixing}

%\end{figure}
In this case, spin exchange energy is decided by the virtual excitations which lead to spin exchange between two sites, and thus depends on the number of orbitals available for the exchange process and the energy of these orbitals. There are multiple pathways available for these spin exchange processes. Two such exchange processes are shown in Fig.~\ref{virtual_excitation}, where we have shown the virtual excitations giving rise to the magnetic interactions between two metal ions. These virtual excitations involve the charge transfer from ligand orbitals to the magnetic ion.  Their contribution to magnetic coupling depends on the energy difference between obitals and the on-site columbic repulsions. In the presence of a drive discussed in Sec. IV, these orbitals of the  ligand ion are modified according to the drive amplitude and frequency. Now, we proceed in the same way as the undriven case, but the orbitals $A$ and $B$ are replaced by the hybrid orbitals:
\begin{equation}
	\begin{split}
		&\Ket{P,n}=\cos\frac{\theta}{2}\Ket{A,n}+\sin\frac{\theta}{2}\Ket{B,n+1},\,\\&\Ket{M,n}=\sin\frac{\theta}{2}\Ket{A,n}-\cos\frac{\theta}{2}\Ket{B,n+1},
	\end{split}
\end{equation}
where $n$ denotes the Floquet index,  $\cos\theta=\frac{\delta}{\sqrt{\delta^{2}+4\Omega^{2}}},\,\sin\theta=-\frac{2\Omega}{\sqrt{\delta^{2}+4\Omega^{2}}}$,
and $\delta=\omega-\omega_0$ is the detuning. If the parameter $eEa/\omega\ll1$ (which is the case here, as $\omega\approx10 eV$ and $eEa\approx1eV$), then the hopping is allowed between orbitals within the same photon sector only in the Floquet picture.  Using this fact, we can calculate the hopping elements between metal sites, and the new orbitals can be expressed as:
\begin{equation}
	t_{P}=\left<A,0|P,0\right>t_{A}+\left<B,0|P,0\right>t_{B}=\cos\frac{\theta}{2}t_{A},
\end{equation}
\begin{equation}
	t_{M}=\left<A,0|M,0\right>t_{A}+\left<B,0|M,0\right>t_{B}=\sin\frac{\theta}{2}t_{A}.
\end{equation}
Now, if $t_B\ne0$, the hopping element between $\ket{P/M,-1}$ and metal sites can still be non-zero as:
\begin{equation}
	\begin{split}
		&\Ket{P,-1}=\cos\frac{\theta}{2}\Ket{A,-1}+\sin\frac{\theta}{2}\Ket{B,0},\,\\&\Ket{M,-1}=\sin\frac{\theta}{2}\Ket{A,-1}-\cos\frac{\theta}{2}\Ket{B,0},
	\end{split}
\end{equation}
with
\begin{equation}
	t_{P1}=\left<A,0|P,-1\right>t_{A}+\left<B,0|P,-1\right>t_{B}=\sin\frac{\theta}{2}t_{B},
\end{equation}
and
\begin{equation}
	t_{M1}=\left<A,0|M,-1\right>t_{A}+\left<B,0|M,-1\right>t_{B}=-\cos\frac{\theta}{2}t_{B}.
\end{equation}
As a result of the drive 
\begin{equation}
	H(t)=\Omega e^{-i\omega t}c\dg_{A\sigma}c_{B\sigma}+\Omega^{*}e^{i\omega t}c\dg_{B\sigma}c_{A\sigma},
	\label{orb_mix_drive}
\end{equation}
the magnetic coupling strength now has contributions from different exchange mechanisms which include virtual excitations via four states, i.e $\Ket{P}$, $\Ket{M}$, $\Ket{P,-1}$, $\Ket{M,-1}$ given by:
\begin{equation}
	J_{ex}=E_0+E_1+E_2
	\label{orb_drive_jex}
\end{equation}
where
%All contributions include
\newcommand{\dde}[1]{\Delta_{#1}}
\newcommand{\al}{\alpha}
\newcommand{\bt}{\beta}
\begin{equation}
	E_{0}=\sum_{\al=P,M,P1,M1}\frac{4t_{\al}^{4}}{\dde{\al}^{2}}\left(\frac{1}{U}+\frac{1}{\dde{\al}}\right)
\end{equation}
\begin{equation}
	\begin{split}
		E_1=\frac{1}{2}\sum_{\bt}\sum_{\al,\al\ne\bt}&\frac{8t_{\al}^{2}t_{\bt}^{2}}{(\dde{\al}+\dde{\bt})\dde{\al}\dde{\bt}}+\frac{4t_{\al}^{2}t_{\bt}^{2}}{(\dde{\al}+\dde{\bt})\dde{\al}^{2}}\\&+\frac{4t_{\al}^{2}t_{\bt}^{2}}{(\dde{\al}+\dde{\bt})\dde{\bt}^{2}}+\frac{8t_{\al}^{2}t_{\bt}^{2}}{U\dde{\al}\dde{\bt}},
	\end{split}
\end{equation}
and 
\begin{equation}
	\begin{split}
		E_2&=\frac{8t_{P}t_{P1}t_{M}t_{M1}}{\dde{P1}U_{m1}\dde{M1}}+\frac{8t_{P}t_{P1}t_{M1}t_{M}}{\dde{P}U_{1}\dde{M}}+\frac{4t_{P}t_{P1}^{2}t_{P}}{\dde{P}U_{1}\dde{P}}\\&+\frac{4t_{M}t_{M1}^{2}t_{M}}{\dde{M}U_{1}\dde{M}}+\frac{4t_{P}^{2}t_{P1}^{2}}{\dde{P1}U_{m1}\dde{P1}}+\frac{4t_{M}^{2}t_{M1}^{2}}{\dde{M1}U_{m1}\dde{M1}},
	\end{split}
\end{equation}
where, $\dde{\al}=U-E_{\al}$, $U_1=U+\omega$, $U_{m1}=U-\omega$  with
\begin{equation}
	E_{P/M}=E_{A}+\frac{\delta}{2}\mp\sqrt{\left(\frac{\delta}{2}\right)^{2}+\Omega^{2}},
\end{equation}
and
\begin{equation}
	E_{P1/M1}=E_{P/M}-\omega
\end{equation}
is the energy of Floquet states $\ket{P/M,-1}$.
\section{ Changes in AFM coupling due to orbital hybridization on each metal site}
\label{metalsite_details}
\begin{figure}[H]
	\centering
	\includegraphics[scale=0.5]{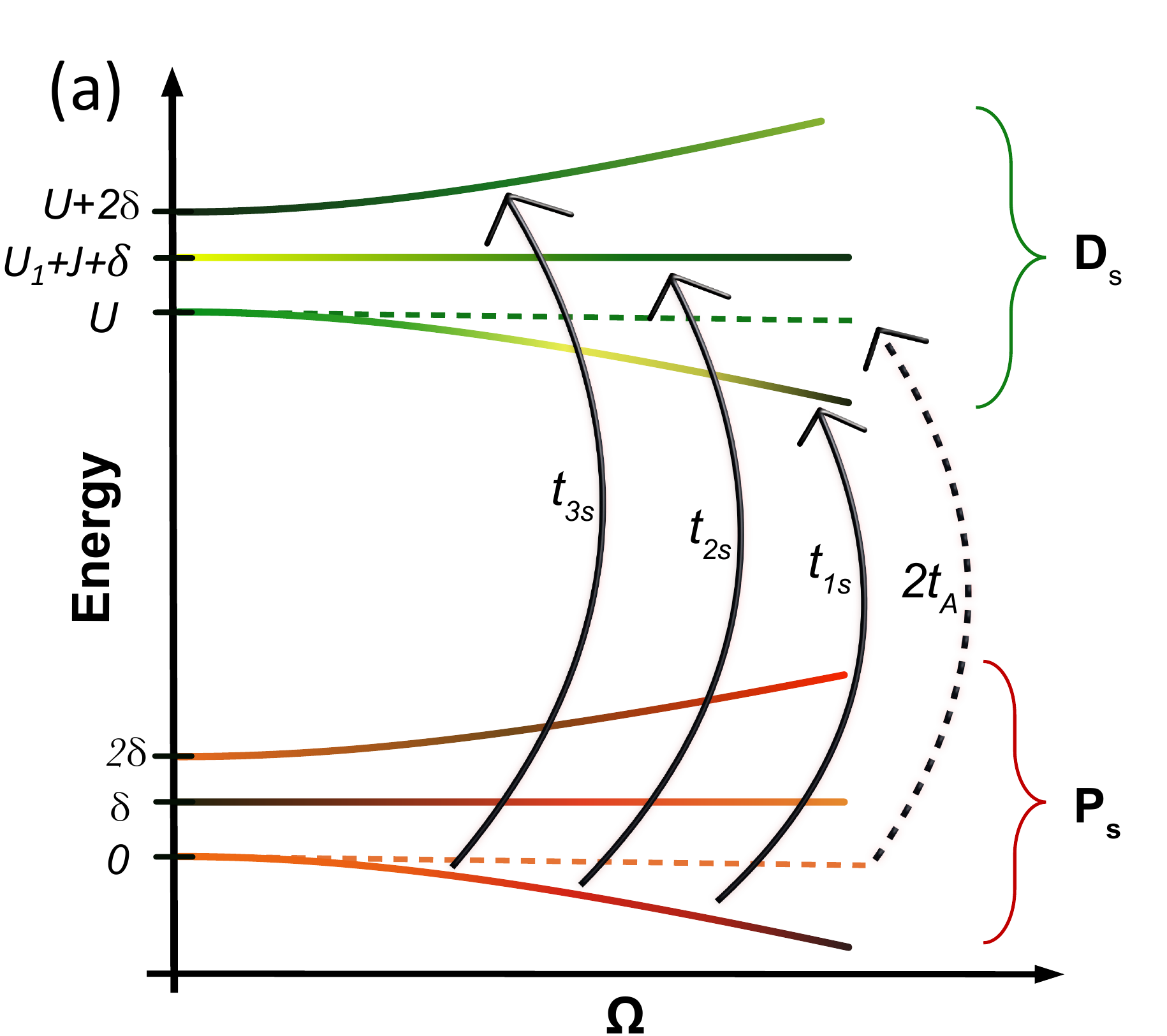}	
	\includegraphics[scale=0.5]{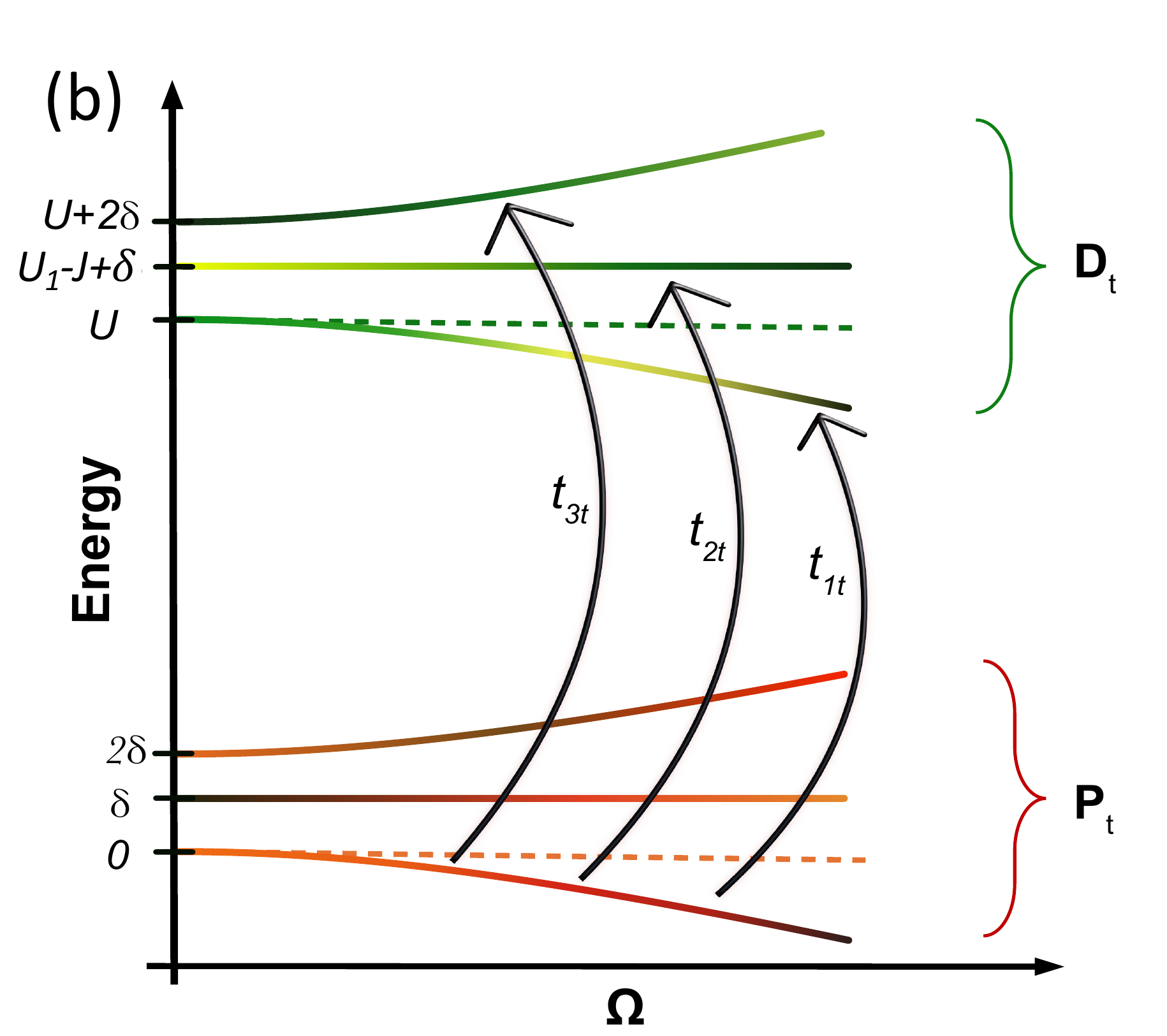}
	\caption{This diagram shows the effect of the periodic drive on the energy levels of a two metal site and two orbital model discussed in Eq.~\ref{Rabi_singlyoccupied} and Eq.~\ref{rabiD}. The lower levels shown in shades of red represent the states connected to the low energy subspace of the undriven model, and the lines in green show the states available for virtual excitations which belong to the doubly occupied sector. These excitations are shown by solid arrows for the driven model, and by the dashed arrow for the undriven case. For clarity, we show the excitations for the singlet (left) and triplet (right) sectors in different diagrams. Here the subspace $P$ and $D$ refers to the singly and doubly occupied states respectively, and the subscript $t/s$ denotes the singlet or triplet nature.  }
	\label{metal_orbital_rabi_splitting}
	
	%Reason why these changes vanish for a special case of ta=tb, tab=0, U1=U-J? For triplet sector hopping to DBB,DAA - same photon number is always zero and for PABt to DABt it is ta-tb when tab=0, i.e all $ti_t=0$, and thus only the stark shift in singlet sector are responsible for any changes - whic are non zero only if U1+J\neU.	
	%Reason why dashed line is shown only in one case - for triplet sector - even when tab is non-zero, the virtual excitation won't belong to the doublon corresponding to two electrons in the same orbital...it will be at U+wo.
\end{figure}

In the undriven model, the spin interactions arise due to virtual excitations between singly and doubly occupied sectors. For large $U,$ the low energy subspace is described by an effective spin hamiltonian. In the presence of a periodic drive which couples two orbital on each metal site, all the states in this subspaces undergo some changes. These changes are best studied using the Floquet formalism, where many singly occupied states now hybridize and the new levels are given by the eigenstates of Floquet hamiltonian. Usually the hopping amplitudes are much smaller in comparison to other energy scales in the problem, and thus we treat the hopping part of the hamiltonian as a perturbation to Floquet hamiltonian obtained from $H_k+H_0+H(t)$. The schematic of the changes in the energy levels of this hamiltonian  is shown in Fig.~\ref{metal_orbital_rabi_splitting} as a function of the drive amplitude for some of the eigenstates relevant for the exchange interactions. 

For the undriven case, there is only one energy eigenstate available for virtual excitations to the doubly occupied sector as denoted by the dashed arrow in Fig.~\ref{metal_orbital_rabi_splitting}. This virtual process lifts the degeneracy between singlet and triplet sectors resulting in a magnetic coupling strength~$\frac{4t_A^2}{U}$. For the driven case, the low energy subspace is replaced by the eigenstates of the Rabi hamiltonian:
\begin{equation}
	H_{R_P}=\left[\begin{array}{ccc}
		0&\Omega\sqrt{2}&0\\
		\Omega\sqrt{2}&\delta&\Omega\sqrt{2}\\
		0&\Omega\sqrt{2}&2\delta\\
	\end{array}	\right]
	\label{Rabi_singlyoccupied}
\end{equation}
with basis vectors given by $\ket{P^{s/t}_{AA},0},\ket{P^{s/t}_{AB},-1},\ket{P^{s/t}_{BB},-2}$ where $P$ denotes the singly occupied sector with subscripts denoting the orbitals on each site, $s/t$ refers to the singlet and triplet sectors, and the integers denote the photon number. For the singly occupied sector, the effects of drive are independent of the spin configuration. This drive also mixes the doubly occupied sectors in a similar manner but in this case the energy levels are given by eigenstates of the following hamiltonian:
\begin{equation}
	H_{R_{D_{s/t}}}=\left[\begin{array}{ccc}
		U&\Omega\sqrt{2}&0\\
		\Omega\sqrt{2}&U_1\pm J+\delta&\Omega\sqrt{2}\\
		0&\Omega\sqrt{2}&U+2\delta\\
	\end{array}\right]
	\label{rabiD}
\end{equation}
%here, delta=w0-w.
which give rise to a different energy shift for the two sectors if $U_1+J\ne U$. In addition to the changes in the energy levels this drive also changes the eigenstates, and thus the hopping parameters are changed accordingly. The hopping processes in the presence of a periodic drive are shown by solid arrows in Fig.~\ref{metal_orbital_rabi_splitting}. In terms of these hopping amplitudes, the new magnetic coupling strength is given by:
\begin{equation}
	E_s-E_t=\sum_{i=1}^3\frac{t_{is}^2}{E_{d_s}^i-E_{p_s}^1}-\sum_i\frac{t_{it}^2}{E_{d_t}^i-E_{p_t}^1}
\end{equation}
where $E_{d_{s/t}}^i $ is the energy corresponding to the eigenstate $\ket{D_{s/t}}$ of hamiltonian $H_{R_{D_{s/t}}} $ in Eq.~\ref{rabiD}, and 
\begin{equation}
	t_{is/t}=\bra{P_{s/t}^1}H_t{\ket{D_{s/t}^i}},
\end{equation} where $\ket{P_{s/t}^1}$ is the eigenstate corresponding to the eigenvalue $\delta\left(1-\sqrt{1+2\left(\frac{\Omega}{\delta}\right)^2}\right)$ of the hamiltonian $H_{R_P}$ for the singly occupied sector, and $H_t$ is the hopping part given by:
\begin{equation}
	\begin{split}
		H_t=-\sum_{\sigma,\alpha=A,B}t_\alpha c_{1\alpha\sigma}\dg c_{2\alpha\sigma}-t_{AB}\sum_{\sigma,i\ne j}c_{1A\sigma}\dg c_{2B\sigma}+\text{h.c},
		\label{metal_site_hopping}
	\end{split}
\end{equation}% of the hamiltonian given by Eq. ~\ref{metal_site_hopping}.
%$H$ in %Eq.~\ref{fullH_metalsite} given by Eq.~\ref{metal_site_hopping}.

% exchange interaction arises as a result of virtual excitations to higher energy subspace which have some doubly occupied sites. These excitations are different for singlet and triplet case and periodic affect these different virtual excitations in a different manner. In order to understand the details of these changes, we treat singlet and triplet sectors separately and see the changes which arise in these different energy sectors as a function of drive parameters. We focus on those states of the singly occupied sectors which are mixed by the drive and  similarly in the doubly occupied subspace, we consider those states which are involved in the virtual excitation processes and are affected by the periodic drive. 

\section{Approximate values of Dipole moment  matrix element using Slater type orbitals}
\label{STO}
\subsection{Dipole moment for ligand orbitals}
The strength of the drive used in the orbital hybridization scheme depends on the value of the dipole moment operator between the two orbitals.  Here, we have used Slater type orbitals~\cite{Slater_PhysRev.36.57} to calculate these dipole moments. The wavefunction of these orbitals is given by:
\begin{equation}
	\ket{\Psi_{n,l,m}(\textbf{r})}=R_n(r)Y_{l}^m(\textbf{r}),
\end{equation}
where 
\begin{equation}
	R_n(r)=(2\zeta)^n\sqrt{\frac{2\zeta}{(2n)!}}r^{n-1}e^{-\zeta r},
\end{equation}
with  $\zeta=\frac{Z^*}{n}$ and $Z^*$ is the effective charge which can be calculated using Slater's rules when distances are expressed in atomic units (1 unit =$a_0$).
Using these Slater type orbitals, we approximate the expectation value of the position operator for different ligand and transition metal orbitals as shown in Table \ref{dipolemoment}. 

\subsection{Dipole moment for \textit{d}-\textit{d} transitions}
In addition to the dipole transitions between different parity orbitals in the ligand ion, we also studied the effects of on-site \textit{d-d} transitions. Although for pure \textit{d} orbitals this kind of transitions are forbidden, but there are many different \textit{d-p} mixing mechanisms available in transition metal compounds which allow these dipole transitions.
In most of the transition metal compounds, $e_g$ and $t_{2g}$ orbitals are not purely of $d$ character but has some contribution from $p$ orbitals. These \textit{p} orbitals can either belong to the same magnetic ion or to the ligand ion.  In the absence of a center of symmetry, the $t_{2g}$ orbitals can mix with $p$ orbitals of the same ion. These $d$ orbitals can also mix with the $p$ orbitals of the ligand ion due to covalency efffects. This kind of mixing is allowed for both tetrahedral and octahedral crystal fields and is one of the most prominent mechanism for \textit{d-p} mixing as indicated by the studies of the intensities of \textit{d-d} observed in many transition metal compounds~\cite{dpnaiman1961interpretation, dpliehr1957intensities,dpballhausen1958intensities,dpballhausen1959intensities}.  In transition metal compounds, the outermost electrons reside in $d$ orbitals and the covalent bonding between the metal and the ligand ion can result in \textit{d-p} mixing, and hence modifying the wavefunction of $d$ orbitals as follows:
\begin{equation}
	\ket{\psi'_{d_i}}=\frac{1}{\sqrt{1+\alpha^2}}\left(\ket{\psi_{d_i}}+\alpha\ket{\chi_{p}}\right),
\end{equation}
where $\chi$ denotes the orbitals of ligand ions and $\alpha\ll1$ (check Ref.~\cite{covalency_PhysRev.130.517} for more details of \textit{d-p} mixing). As a result,  the dipole moment operator $e\left|\left<\psi'_{t_{2g}}|\textbf{r}|\psi_{e_g}\right>\right|\approx e\left|\frac{\alpha}{\sqrt{1+\alpha^2}}\left<\psi_{3d_{z^2-r^2}}|\textbf{r}|\chi_{p_z}\right>\right|$ depends on the arrangement of ligand ions around the metal ion. This quantity can be estimated from Slater like orbitals if the mixing parameter is known which is usually difficult to determine. Since, this dipole moment is also proportional to the oscillator strength which can be calculated directly from the absorption spectra of these complexes. In some tetrahedral complex salts~\cite{dpballhausen1958intensities}, the dipole moment between two different $d$ orbitals belonging to $t_{2g}$ and $e_g$ sets can be as high as $0.5$ Debye$ =0.1e\AA$. This kind of \textit{d-d} transition also occur in some transition metal trichalcogenides like $\text{NiPS}_3$ ~\cite{dpPhysRevLett.120.136402} but the associated dipole moment would be much smaller as indicated by the extremely weak absorption for this peak.
%In this paper, authors show that covalency factor $\alpha\approx0.6$ - and for tert
%Within this formalism, we can approximate the expectation value of the position operator $\left<\psi_{e_g}|\textbf{r}|\psi'_{t_{2g}}\right>$. For e.g, in tetrahedral geometry with metal ligand bond length M-L$\approx 2\AA$, the quantity $\left|\left<\psi_{3d_{z^2}}|\textbf{r}|\chi_{3p_z}\right>\right|\approx 0.1\AA,\,\left|\left<\psi_{3d_{xy}}|\textbf{r}|\chi_{3p_z}\right>\right|\approx 0.05\AA$ and for most of the cases $\alpha<0.1$, which gives $\left|\left<\psi_{t_2g}'|\textbf{r}|\psi_{e_g}'\right>\right|<0.01\AA$.
%mathematica file - Slater orbital -t2g-(dxy,dyz,dxz)-eg-(dx2y2,dz2)
\begin{table}[H]
	\centering
	\begin{tabular}{|c|c|c|c|}
		
		\hline
		Element&$A$&$B$&$\left<\psi_A|\textbf{r}|\psi_B\right>$ 
		\tabularnewline
		\hline O&$2s$&$2p_z$&$0.6\,a_0\,\hat{z}$
		\tabularnewline
		\hline S&$3s$&$3p_z$&$1.1\,a_0\,\hat{z}$
		\tabularnewline
		\hline Sc&$3d_{z^2-r^2}$&$3p_z$&$0.3\,a_0\,\hat{z}$
		\tabularnewline
		\hline Mn&$3d_{z^2-r^2}$&$4p_z$&$1.0\,a_0\,\hat{z}$
		\tabularnewline
		\hline
	\end{tabular}
	\caption{Position operator matrix elements between different orbitals calculated using Slater type orbitals.}
	\label{dipolemoment}
\end{table}

\section{Vibronic coupling estimate}
\begin{figure}
	\includegraphics[scale=0.5]{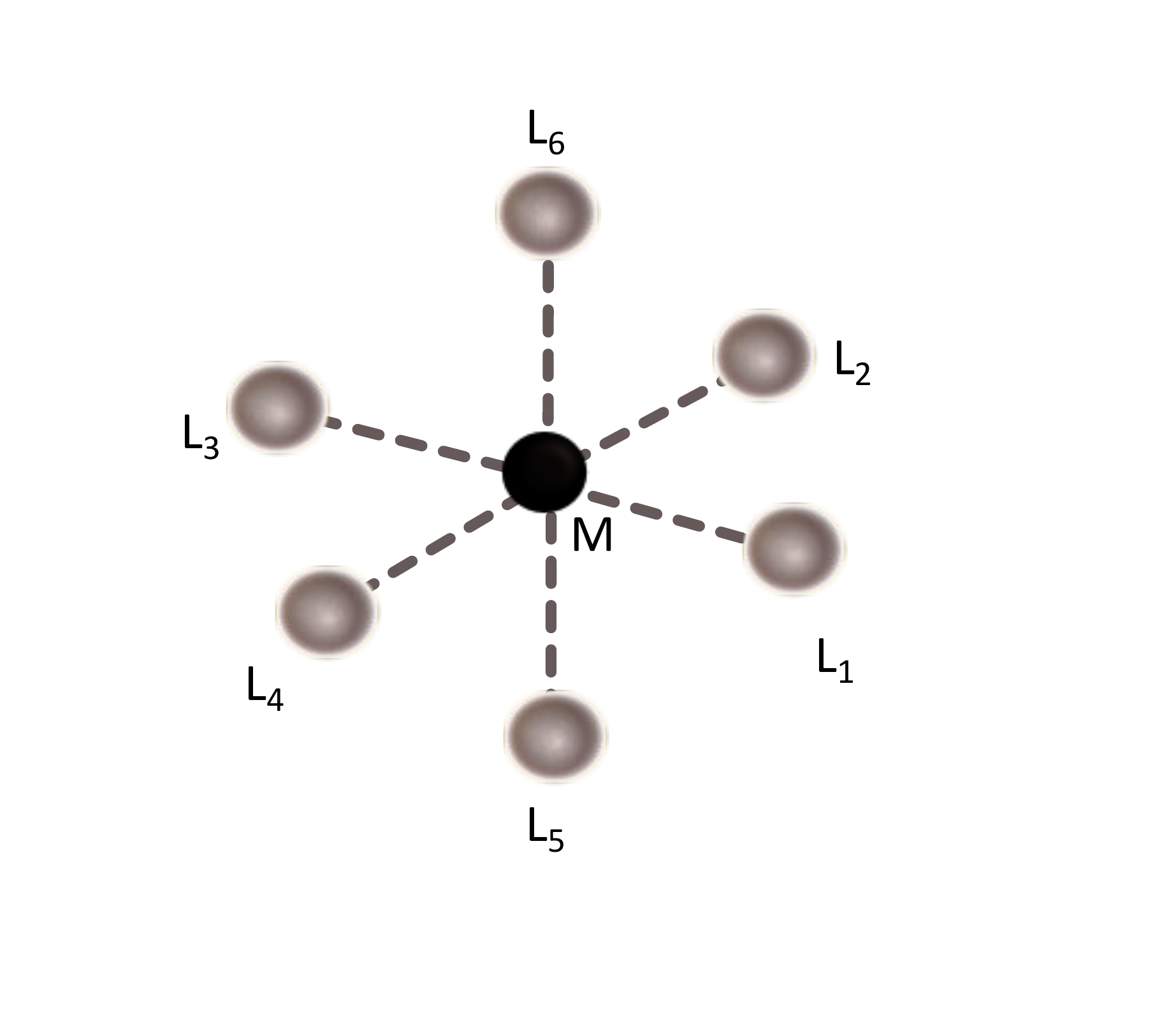}
	\caption{Arrangement of ligand ions around a transition metal ion in octahedral geometry.}
	\label{ligandarrangement}
\end{figure}

In this section, we calculate the matrix element between two $d$ orbitals for a phonon drive. We assume that the transition metal ion is surrounded by an octahedral arrangement of ligand ions, and the phonon mode involves the symmetric motion of ligand ions perpendicular to the metal ligand bond.  For the geometry shown in Fig. \ref{ligandarrangement}, the potential around TM ion due to ligand ions is given by
\begin{equation}
	V(\textbf{r},t)=\sum_{L=1}^8 \frac{q_L e^2}{4\pi\epsilon_0|\textbf{a}_L(t)-\textbf{r}|}
\end{equation}
where $q_L$ is the charge on ligand ion (in units of $e$), $\textbf{a}_L$ is the position vector of  ligand $L$ from the center of the TM ion, and 
\begin{equation}
	\textbf{a}_L(t)=\textbf{a}^0_L+\textbf{u}_L(t)
\end{equation}
where $\textbf{a}^0_L$ is the equilibrium distance of M-L bond, and $\textbf{u}_L$ is the phonon amplitude. For small $\textbf{u}_L$, we can expand $V$ around its equilibrium value as follows:
\begin{equation}
	\begin{split}
		V(\textbf{r},t)=\frac{q_Le^2}{4\pi\epsilon_0}\sum_{L=1}^8 \left(\frac{1}{|\textbf{a}^0_L-\textbf{r}|}-\frac{(\textbf{a}^0_L-\textbf{r})\cdot \textbf{u}_L(t)}{|\textbf{a}^0_L-\textbf{r}|^3}+...\right),
	\end{split}
\end{equation}
and thus upto first-order, the perturbation is given by:
\begin{equation}
	H'\approx-\frac{q_Le^2}{4\pi\epsilon_0}\sum_j^8\frac{(\textbf{a}^0_L-\textbf{r})\cdot\textbf{u}_L(t)}{|(a_L^0)^2+r^2|^\frac{3}{2}}\left(1+3\frac{\textbf{a}_L^0\cdot \textbf{r}}{|(a_L^0)^2+r^2|}+... \right).
\end{equation}
Now, the only terms which can couple two $d$ orbitals are:
\begin{equation}
	\left<d_\alpha|H'|d_\beta\right>=\frac{3q_Le^2}{4\pi\epsilon_0}\sum_{L=1}^8\left<d_\alpha\left|\frac{(\textbf{r}\cdot \textbf{u}_L(t))(\textbf{a}_L^0\cdot \textbf{r})}{|(a_L^0)^2+r^2|^\frac{5}{2}}\right|d_\beta\right>.
\end{equation}
Furthermore, the matrix element $\left<d_\alpha|\textbf{r}_k\textbf{r}_l|d_\beta\right>$ is non-zero for the  cases shown in Table \ref{dxy}.
\begin{table}[H]
	\centering
	\begin{tabular}{|c|c|c|c|}
		
		\hline
		$d_\alpha$&$d_\beta$&$\textbf{r}_k\textbf{r}_l$&$\left<d_\alpha|\textbf{r}_k\textbf{r}_l|d_\beta\right>(\AA^2)$
		\tabularnewline
		\hline
		
		\hline $d_{xy}$&$d_{yz}$&$xz$&$0.4$
		\tabularnewline
		\hline $d_{xy}$&$d_{xz}$&$yz$&$0.4$
		\tabularnewline
		\hline$d_{yz}$ &$d_{xz}$&$xy$&$0.4$
		\tabularnewline
		\hline$d_{yz}$ &$d_{x^2-y^2}$&$yz$&$0.4$
		\tabularnewline
		\hline$d_{xz}$ &$d_{x^2-y^2}$&$xz$&$0.4$
		\tabularnewline
		\hline$d_{xy}$ &$d_{z^2-r^2}$&$xy$&$0.4$
		\tabularnewline
		\hline$d_{yz}$ &$d_{z^2-r^2}$&$yz$&$0.2$
		\tabularnewline
		\hline$d_{xz}$ &$d_{z^2-r^2}$&$xz$&$0.2$
		\tabularnewline
		\hline
		
	\end{tabular}
	\caption{Matrix element $\left<\textbf{r}_k\textbf{r}_l\right>$ between two $3d$ orbitals of Ti  calculated using Slater type orbitals.}
	\label{dxy}
\end{table} 
In $\text{RTiO}_3$, the  ligand-metal distance, $|a_L^0|\approx2\AA$, and thus the matrix element coupling two $d$ orbitals  becomes:
\begin{equation}
	|\left<d_{\alpha}\left| H'\right|d_{\beta}\right>|\approx0.5u(t) \text{eV}=0.25(e^{i\omega t}+e^{-i\omega t})u_0 \text{eV}
\end{equation}
for a $g$ symmetry phonon mode, where $u_0$ is the displacement (in units of $\AA$) of the ligand ion perpendicular to the M-L bond.

%mention some example 
%covalency induces mixing between metal and ligand orbitals. #effects of covalency can be taken into account by considering the following form of orbital wavefunctions
%\bibliographystyle{unsrt}
\bibliographystyle{apsrev4-1}   %use this part if you are using revtex4-1
\bibliography{magnetismref.bib}
\end{document}